\documentclass[aps, prx, superscriptaddress, twocolumn, amssymb, longbibliography]{revtex4-2}

\usepackage{mathtools} 
\usepackage{graphicx} 
\usepackage{placeins} 
\usepackage{booktabs} 

\begin{document}


\title{Generation of Ultrahigh Anomalous Hall Conductivities \\ via Optimally Prepared Topological Floquet States}

\author{Andrew Cupo}
\email{a.cupo@northeastern.edu}
\affiliation{Department of Physics, Northeastern University, Boston, Massachusetts, 02115, USA}
\affiliation{Department of Physics and Astronomy, Dartmouth College, Hanover, New Hampshire, 03755, USA}

\author{Hai-Ping Cheng}
\email{ha.cheng@northeastern.edu}
\affiliation{Department of Physics, Northeastern University, Boston, Massachusetts, 02115, USA}

\author{Chandrasekhar Ramanathan}
\email{chandrasekhar.ramanathan@dartmouth.edu}
\affiliation{Department of Physics and Astronomy, Dartmouth College, Hanover, New Hampshire, 03755, USA}

\author{Lorenza Viola}
\email{lorenza.viola@dartmouth.edu}
\affiliation{Department of Physics and Astronomy, Dartmouth College, Hanover, New Hampshire, 03755, USA}


\begin{abstract}
Ultrafast quantum matter experiments have validated predictions from Floquet theory -- notably, the dynamical modification of the electronic band structure and the light-induced anomalous Hall effect, via monotonic modulation of the driving amplitude. Here, we demonstrate how new physics is uncovered by leveraging quantum optimal control techniques to design Floquet amplitude modulation profiles. We discover a fundamentally different regime of topological transport, whereby the optimal oscillatory preparation protocol functions as a non-adiabatic topological pump: as a result, ultrahigh time-averaged anomalous Hall conductivities emerge, that reach up to around seventy times the values one would expect from the Chern number of the targeted Floquet state. The optimal protocols achieve $>$99\% fidelity at the topological energy gap closing point -- a twenty-fold improvement over standard monotonic approaches in as little as ten Floquet cycles -- while unexpectedly generating the predicted ultrahigh conductivities. Our findings demonstrate that optimally prepared non-equilibrium quantum states can access transport regimes not achievable in the corresponding equilibrium system or even by applying conventional Floquet approaches, opening new avenues for ultrafast quantum technologies and topological device applications.
\end{abstract}


\maketitle




\section{Introduction}
\label{sec:intro}

Time-periodic modulation of a quantum system results in an effective renormalization of observable properties of interest, with Floquet theory being a powerful mathematical framework for modeling and prediction \cite{goldman2015, bukov2015universal}. Motivated by continuous progress in experimental coherent and ultrafast control capabilities, Floquet modulation schemes have attracted increasing attention as a means for accessing novel states of matter and non-equilibrium phenomena \cite{basov2017towards, oka2019floquet, giovannini2020floquet, harper2020topology, rudner2020band, rodriguez2021low}. An especially prominent example is the dynamical modulation and detection of the electronic band structure of quantum materials by means of time- and angle-resolved photoemission spectroscopy (TR-ARPES) \cite{wang2013observation, mahmood2016selective, reimann2018subcycle, reutzel2020coherent, keunecke2020electromagnetic, aeschlimann2021survival, zhou2023pseudospin, zhou2023floquet, bao2024manipulating, bielinski2025floquet}. In the important case of graphene \cite{choi2025observation, merboldt2025observation}, time-reversal symmetry-breaking circularly polarized radiation opens a topological energy gap at the $K$ and $K'$ points. This leads to a non-zero (approximately) quantized time-averaged anomalous Hall conductivity, which has been predicted theoretically \cite{oka2009photovoltaic, sato2019microscopic} and measured experimentally \cite{mciver2020light}. Floquet topological insulators have likewise been broadly discussed in recent literature \cite{lindner2011floquet, cayssol2013floquet, rudner2013anomalous, roy2017periodic, rudner2020band}.

Notwithstanding the above advances, challenges remain in realizing the full potential of the periodic driving in experiment. For an ideal drive with infinite time-periodicity, Floquet's theorem guarantees that the evolution may be exactly described in terms of an effective time-independent Floquet Hamiltonian \cite{goldman2015, bukov2015universal}. Since real experiments have finite duration, however, addressing how the periodic driving is ``launched'' via amplitude modulation becomes essential. In typical experiments \cite{wang2013observation, mciver2020light}, the duration of the high-intensity laser pulse is only on the order of one picosecond to prevent material damage. A question then arises: When the amplitude of the external control is modulated from zero up to a targeted value, how close is the final quantum state to the one that would be predicted from Floquet theory? The \textit{preparation of Floquet states} has been considered in the literature from different perspectives \cite{d2015dynamical, seetharam2015controlled, ho2016quasi, ge2017topological, desbuquois2017controlling, bukov2018reinforcement, crowley2019topological, bandyopadhyay2020dissipative, ge2021universal, minguzzi2022topological, castro2022floquet, lucchini2022controlling, ito2023build, schnell2024dissipative, schindler2024counterdiabatic, ritter2025autonomous}; a few important points are worth highlighting in this regard. 

Given advances in ultrafast spectroscopic methods, it is now possible to achieve sub-cycle resolution in experiments, as demonstrated for both atoms and solids. For instance, for Ne atoms Floquet states are established within ten driving cycles via observation of the relative sideband weighting from photoelectron spectroscopy \cite{lucchini2022controlling}. Additionally, in the case of the topological surface states of Bi$_2$Te$_3$, Floquet sidebands emerge in TR-ARPES signatures within a single optical driving cycle \cite{ito2023build}.

\begin{figure*}[!htb]
\centering{\includegraphics[scale=0.7]{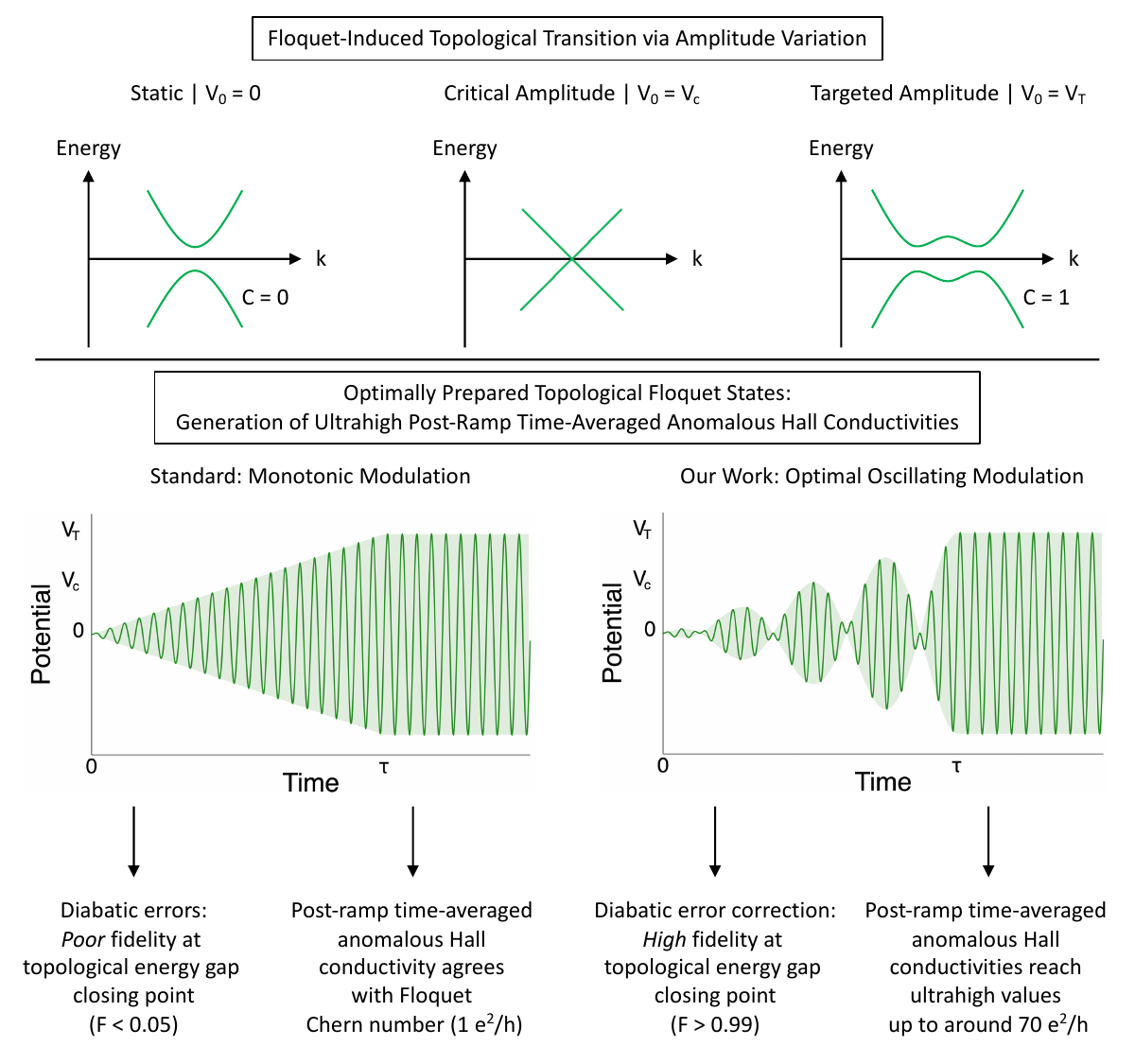}}
\caption{
{\bf Qualitative overview of key findings in this paper.} 
[Top] Demonstration of a topological transition induced by varying the amplitude of a Floquet drive in 2D quantum matter, as evidenced by the change in Chern number across the (quasi)energy gap closing point. 
[Bottom Left] Amplitude modulation from zero through the critical amplitude up to a targeted value \textit{monotonically}. As one expects, monotonic passage produces diabatic errors, resulting in a poor Floquet state preparation fidelity at the topological energy gap closing point. The post-ramp time-averaged anomalous Hall conductivity is still consistent with the Floquet Chern number of the lower targeted band. 
[Bottom Right] The optimal oscillatory amplitude modulation results in high Floquet state preparation fidelities at the topological energy gap closing point via diabatic error cancellation. Most importantly, the oscillatory protocol functions as a non-adiabatic topological pump in 2D that results in ultrahigh post-ramp time-averaged anomalous Hall conductivities.
}
\label{fig:overview}
\end{figure*}

A key challenge arises, however, when the target Floquet state belongs to a different universality class of the original undriven many-body state, as the system must necessarily undergo a quantum phase transition associated with a gap closing \cite{d2015dynamical, ho2016quasi, ge2017topological, desbuquois2017controlling, crowley2019topological, ge2021universal, minguzzi2022topological}. Ramping the amplitude of the Floquet drive then inevitably leads to energy level crossings in real time and consequent preparation error. In the simplest case, the use of \textit{monotonic} linear \cite{d2015dynamical, ge2017topological, ge2021universal} and (analytically optimized) exponential \cite{ho2016quasi} ramps leads to significant fidelity limitations near the crystal momentum where the energy gap closes. Nevertheless, for two-dimensional (2D) systems, if the amplitude is modulated through a critical value that would in principle lead to a topological transition in the Floquet system, topologically protected edge states still emerge when the spatial boundaries are opened \cite{d2015dynamical, ge2017topological, ge2021universal}. Furthermore, under appropriate conditions, the post-preparation time-averaged anomalous Hall conductivity is close to the quantized value expected from a linear response theory of electronic transport applied to the Floquet system \cite{sato2019microscopic, sato2019light, ge2021universal}. This feature -- that the topologically protected edge states and quantized anomalous Hall conductivities emerge despite the poor fidelity at the topological gap closing point -- leads naturally to the question: If one devises a method for improving the fidelity, will there be any substantial changes to relevant physical observables?

How to improve the fidelity when preparing Floquet states if energy gap closings are involved has been addressed via several methodologies \cite{desbuquois2017controlling, bukov2018reinforcement, crowley2019topological, minguzzi2022topological, castro2022floquet, schindler2024counterdiabatic}. Towards general applicability, a few works have approached this problem as an optimization of the parameterized external control: specifically, this has been explored from the perspective of reinforcement learning \cite{bukov2018reinforcement}, quantum optimal control (QOC) \cite{castro2022floquet}, and counterdiabatic driving \cite{schindler2024counterdiabatic}. Ultimately, the most appropriate framework depends on the system of interest and the form of the external control(s) that can be implemented in practice. For laser driving of the electronic degrees of freedom in quantum materials, the QOC approach seems especially compelling, in that the amplitude modulation turn-on profile of the Floquet drive can be adjusted systematically to maximize the fidelity for a given preparation time, subject to implementation constraints \cite{castro2022floquet}. With these developments in mind, the remaining essential question is: When the fidelity is enhanced by these means, what is the corresponding effect on physical observables of interest, in particular on the electronic transport properties?

Motivated by the above, in this paper we consider the ultrafast preparation of topologically non-trivial Floquet states to high fidelity, starting from topologically trivial states. Utilizing the QOC formalism, oscillating amplitude modulation profiles are designed to improve the prepared state fidelity from less than 0.05 to better than 0.99 in as little as ten Floquet driving cycles. Repeated traversals through the quantum critical point produce interference effects in the context of Landau-Zener-St\"uckelberg physics. Physically, our key finding is that the optimal oscillatory preparation protocols function as \textit{non-adiabatic topological pumps}, that lead to post-ramp time-averaged anomalous Hall conductivities reaching ultrahigh values up to around 70 $e^2/h$, even though the Chern number of the corresponding targeted Floquet states is only 1 in our model. Fig.~\ref{fig:overview} illustrates how these optimized protocols simultaneously achieve high fidelity state preparation and generate anomalous Hall conductivities that dramatically exceed expectations. 

The outline of the paper is as follows. We first demonstrate our approach using the 2D quantum well, establishing a useful set of static and Floquet parameters such that variation of the driving amplitude leads to a topological phase transition (Sec.~\ref{sec:qw}). With the preliminaries laid out, we describe the theoretical framework for the quantum dynamics of preparing Floquet states via amplitude modulation of the time-periodic driving term (Sec.~\ref{sec:preparation}). As a reference, we explore the Floquet state preparation fidelity of both elementary monotonic (Sec.~\ref{sec:monotonic}) and simple oscillating (Sec.~\ref{sec:oscillating}) amplitude modulation profiles (ramps). On the basis of QOC methods, we then design ramps that maximize the fidelity at the topological energy gap closing point, while keeping the maximum frequency component as small as possible (Sec.~\ref{sec:qoc}). Afterwards we simulate the time-dependent anomalous Hall transport properties and determine their post-ramp time-averages under the action of the different ramping schemes (Sec.~\ref{sec:transport}): The oscillatory preparation protocols function as non-adiabatic topological pumps, leading to ultrahigh time-averaged anomalous Hall conductivities. Towards practicality, we further assess how robust the fidelities and time-averaged anomalous Hall conductivities from the optimally designed ramps are against random noise of varying strengths (Sec.~\ref{sec:robustness}). We wrap up with a summary and outlook with possibilities for future research (Sec.~\ref{sec:conc}).


\section{Floquet-Induced Topological Transition in a 2D Quantum Well}
\label{sec:qw}

For illustration purposes, within this work we consider the static system to be a two-band quantum well model \cite{bernevig2006quantum, lindner2011floquet, kumar2020linear}. Physically, the Hamiltonian describes the behavior of electrons in a 2D layer which is sandwiched between two bulk materials. Mathematically, let $\vec{\sigma}$ denote the vector of Pauli spin matrices; we then have
\begin{equation}
H_{0, \boldsymbol k}
=
\vec{d}_{\boldsymbol k}
\cdot
\vec{\sigma}
,
\label{ham_static}
\end{equation}
where ${\boldsymbol k}\equiv (k_x,k_y)$ is the 2D wave-vector in the first Brillouin zone and $\vec{d}_{\boldsymbol k}$ represents an effective spin-orbit field,
\begin{align}
\vec{d}_{\boldsymbol k}
\equiv 
&\,A \,\text{sin}(k_x l_x)
\hat{x}
+
A \,\text{sin}(k_y l_y)
\hat{y}
\notag
\\
&+
[M - 4B + 2B\, \text{cos}(k_x l_x) + 2B \,\text{cos}(k_y l_y)]
\hat{z}
,
\label{d_vector}
\end{align}
where $A$, $B$, and $M$ are adjustable parameters specific to the combination of materials. The time-independent Schr\"{o}dinger equation reads
\begin{equation}
H_{0, \boldsymbol k} \varphi_{n \boldsymbol k}
=
E_{n \boldsymbol k} \varphi_{n \boldsymbol k}
.
\label{tise}
\end{equation}

A time-periodic control of the form \cite{cayssol2013floquet}
\begin{equation}
D(t)
\equiv 
V_0
[
(\sigma_x - i \sigma_y) e^{i \Omega t}
+
(\sigma_x + i \sigma_y) e^{-i \Omega t}
]
\label{floquet_drive}
\end{equation}
can be applied to the quantum well, yielding the time-dependent Hamiltonian
\begin{equation}
H_{\boldsymbol k}(t) 
=
H_{0, \boldsymbol k}
+
D(t)
\label{td_ham_floquet}
\end{equation}
with periodicity $T\equiv 2 \pi / \Omega$:
\begin{equation}
D(t+mT) = D(t) 
\rightarrow 
H_{\boldsymbol k}(t+mT) = H_{\boldsymbol k}(t),
\;\,m\in {\mathbb Z}
.
\label{time_periodic}
\end{equation}
In this case, the adjustable control parameters are the angular frequency $\Omega$ and the potential amplitude $V_0$. In the context of the quantum well, the form of $D(t)$ in Eq.~\eqref{floquet_drive} will correspond to applying circularly polarized radiation \cite{cayssol2013floquet}. We shall show numerically that the time-reversal symmetry-breaking feature of such a drive enables a topological transition when $\Omega$ is fixed and $V_0$ is varied suitably. Additional insights into the chosen form of the drive in Eq.~\eqref{floquet_drive} are provided in Appendix~\ref{sec:app-hf}.

With a time-periodic Hamiltonian, the solutions of the time-dependent Schr\"{o}dinger equation 
\begin{equation}
i \hbar \partial_t \psi_{n \boldsymbol k}(t) 
= 
H_{\boldsymbol k}(t) \psi_{n \boldsymbol k}(t)
\label{tdse}
\end{equation}
can be described in terms of basis states that obey Floquet's factorization Ansatz \cite{floquet1883equations, shirley1965solution, sambe1973steady, rudner2020floquet}:
\begin{equation}
\psi_{n \boldsymbol k}(t) 
\equiv 
e^{-i \epsilon_{n \boldsymbol k} t/\hbar} 
\Phi_{n \boldsymbol k}(t)
, 
\quad
\Phi_{n \boldsymbol k}(t+T) = \Phi_{n \boldsymbol k}(t)
,
\label{floquet_ansatz}
\end{equation}
where the Floquet eigenstates $\Phi_{n \boldsymbol k}(t)$ are $T$-periodic and the Floquet quasienergies $\epsilon_{n \boldsymbol k}$ are time-independent but defined only up to multiples of $m \hbar \Omega$, with $m \in {\mathbb Z}$ labeling different Floquet Brillouin zones. Both $\epsilon_{n \boldsymbol k}$ and $\Phi_{n \boldsymbol k}(t)$ can be determined numerically by means of the Fourier expansion technique \cite{sambe1973steady, rudner2020floquet}, which we have described and used extensively in previous work \cite{cupo2021floquet, cupo2023optical, cupo2025floquet}. Likewise, the electronic band structure, topology, and optical conductivity of the Floquet 2D quantum well have been studied in previous works \cite{lindner2011floquet, kumar2020linear}.

\begin{figure*}[!htb]
\centering{\includegraphics[scale=0.55]{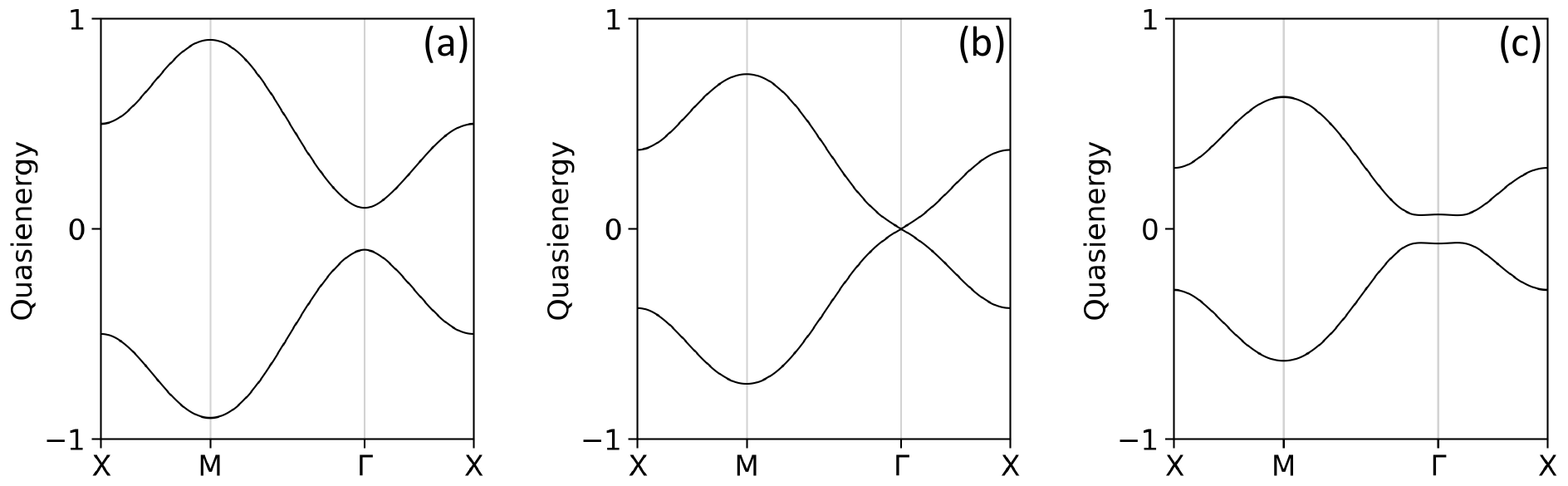}}
\caption{
{\bf Quasienergy band structures of the Floquet driven 2D quantum well.} 
The Floquet photon energy is $\hbar \Omega = 4$ and each panel is the result for a different driving amplitude $V_0$. (a) $V_0 = 0$ corresponds to the limiting case of the static system, with parameters $A = -0.1$, $B = -0.1$, and $M = 0.1$. The computed Chern number of the lower band is $C=0$, indicating that the system is topologically trivial. (b) $V_0 = V_c = 0.31$ is the critical amplitude that leads to the quasienergy gap closing at a crystal momentum of $\boldsymbol k = \Gamma$. (c) $V_0 = V_T = 0.41$ is the driving amplitude of the Floquet state that we attempt to target with subsequent exact quantum dynamics. The quasienergy gap is re-opened with the characteristic appearance of band inversion. The computed Chern number of the lower band is now $C=1$, indicating that a topological transition has been induced.
}
\label{fig:bands}
\end{figure*}

Specifically, to assess the topology of the non-equilibrium states, the concept of the Chern number has been generalized to Floquet systems \cite{dehghani2015out, chen2018floquet, kumar2020linear} by letting 
\begin{align}
C_n
\equiv 
&\frac{i}{2 \pi}
\int_{\textrm{BZ}}^{} d^2 k
\bigg[
\frac{1}{T}
\int_{0}^{T} dt
\nonumber
\\
&\bigg(
\frac{\partial \Phi^\dagger_{n \boldsymbol k}(t)}{\partial k_x} \frac{\partial \Phi_{n \boldsymbol k}(t)}{\partial k_y}
-
\frac{\partial \Phi^\dagger_{n \boldsymbol k}(t)}{\partial k_y} \frac{\partial \Phi_{n \boldsymbol k}(t)}{\partial k_x}
\bigg)
\bigg]
.
\label{chern_number}
\end{align}
On the basis of linear response theory, a Floquet analog of the Thouless-Kohmoto-Nightingale-den Nijs (TKNN) relation has also been found to hold \cite{dehghani2015out, chen2018floquet, kumar2020linear}:
\begin{equation}
\sigma_0^{(xy)}
=
\frac{e^2}{h}
\sum_{n}^{}
f_{n}
C_n
,
\label{floquet_tknn}
\end{equation}
where $f_{n}$ is the occupation of each Floquet mode. Thus, the Chern numbers are connected to the \textit{static} (homodyne) part of the DC Hall conductivity.

For our analysis, we find a useful set of parameters to be: $A = -0.1$, $B = -0.1$, and $M = 0.1$, with unity lattice spacings ($l_x = l_y = 1$). Note that we use dimensionless units ($\hbar = 1$ and $|e| = 1$), although we write fundamental constants explicitly throughout, to enable general applicability. The quasienergy band structures (Eq.~\eqref{floquet_ansatz}) numerically obtained for three different Floquet driving amplitudes (Eq.~\eqref{floquet_drive}) are shown in Fig.~\ref{fig:bands}, where we use $\hbar \Omega = 4$. In panel (a), $V_0 = 0$ provides the static reference, where the bands feature a direct band gap at $\Gamma$. At a critical amplitude of $V_0 = V_c = 0.31$, in panel (b) we see that the gap between the bands closes at $\Gamma$, with an accompanying linear dispersion. At the highest \textit{targeted} amplitude of $V_0 = V_T = 0.41$, the gap re-opens with a visual appearance of band inversion. 

Importantly, note that we are plotting the \textit{unfolded quasienergies} \cite{rudner2020floquet}, which are defined by
\begin{equation}
\lim_{V_0 \to 0} 
\epsilon_{n \boldsymbol k}
=
E_{n \boldsymbol k}
.
\label{unfolded}
\end{equation}
That is, the unfolded quasienergies are those that produce the static energy eigenvalues (Eq.~\eqref{tise}) in the limit where the Floquet amplitude is taken to 0. In our case, since the photon energy ($\hbar \Omega = 4$) is considerably larger than the range of the static energy bands ($< 2)$, the unfolded quasienergies trivially occur in the first Floquet Brillouin zone, with boundaries at $-\hbar \Omega / 2$ and $\hbar \Omega / 2$. We explain these subtleties of Floquet theory in detail in our previous work \cite{cupo2021floquet, cupo2023optical, cupo2025floquet}, see in particular Sec.~III of \cite{cupo2023optical}.

In Fig.~\ref{fig:bands}, the computed Chern numbers of the lower bands in panels (a) and (c) are 0 and 1, respectively. Therefore, we see that variation of the Floquet amplitude for a fixed photon energy results in a topological transition. If the lower band is fully occupied in panels (a) and (c) of Fig.~\ref{fig:bands}, then the Floquet generalization of the TKNN relation (Eq.~\eqref{floquet_tknn}) implies static Hall conductivities of $\sigma_0^{(xy)}
= 0$ and $\sigma_0^{(xy)} = e^2/h$, respectively.


\section{Floquet State Preparation Problem: Exact Quantum Dynamics}
\label{sec:preparation}

As we discussed in the introduction, in order to make the ideal Floquet theory for an infinite time-periodic drive useful in experiment, it is essential to address how the periodicity is established, by turning on the drive. Practically, a system begins at equilibrium and the external control must be amplitude modulated from zero up to a targeted value, after which the drive may remain ``locally'' time-periodic and the properties of interest are measured -- before being eventually ramped back off to prevent detrimental heating effects \footnote{As a point of reference, in the experiments by Wang {\em et al}. \cite{wang2013observation}, the entire duration of the pump laser pulse was approximately 30 cycles of the mid-infrared source.}. We now describe the mathematical formalism required to model this more realistic scenario \cite{d2015dynamical, ge2017topological, ge2021universal, sato2019microscopic, sato2019light}. The time-dependent Hamiltonian is updated to
\begin{equation}
H_{\boldsymbol k}(t) 
=
H_{0, \boldsymbol k}
+
V(t)
=
H_{0, \boldsymbol k}
+
\mathcal{R}(t) D(t)
,
\label{td_ham_general}
\end{equation}
which now includes $\mathcal{R}(t)$ to represent the amplitude modulation:
\begin{equation}
\mathcal{R}(t) 
= 
\begin{cases}
0 & t < t_i \\
R(t) & t_i \leq t \leq t_i + \tau \\
1 & t_i + \tau < t
\end{cases}
,
\label{ramp_general}
\end{equation}
where $t_i$ and $\tau$ denote the initial time and the ramping duration, respectively, and $R(t)$ is the ramping function, subject to the boundary conditions $R(t_i)=0$ and $R(t_i + \tau)=1$. Note that, in practice, we set $t_i = 0$ in the numerical simulations for convenience. To accommodate a general occupation profile across energy eigenstates, it is necessary to describe the quantum state in terms of a density operator. Specifically, we take the initial state to have the form
\begin{equation}
\rho_{\boldsymbol k}(t_i)
=
\sum_{n}^{}
f_{n \boldsymbol k} \,
\varphi_{n \boldsymbol k}
\varphi^{\dagger}_{n \boldsymbol k}
,
\label{initial_density_matrix}
\end{equation}
indicating that the eigenstates defined by Eq.~\eqref{tise} each have the occupation $0 \leq f_{n \boldsymbol k} \leq 1$. Note the normalization condition $\textrm{Tr}\{\rho_{\boldsymbol k}(t_i)\} = N_p$ ($\forall \boldsymbol k$), where $N_p$ is the total number of particles in the system. The time-evolution of this initial density operator is governed by the Liouville-von Neumann equation for unitary quantum dynamics
\begin{equation}
i \hbar \partial_t \rho_{\boldsymbol k}(t) 
= 
[ H_{\boldsymbol k}(t) , \rho_{\boldsymbol k}(t) ]
,
\label{lvn_equation}
\end{equation}
where the Hamiltonian is the one specified in Eq.~\eqref{td_ham_general}. The exact solution reads 
\begin{equation}
\rho_{\boldsymbol k}(t)
=
U_{\boldsymbol k}(t,t_i)
\rho_{\boldsymbol k}(t_i)
U^{\dagger}_{\boldsymbol k}(t,t_i)
\label{lvn_solution}
\end{equation}
in terms of the time-evolution operator
\begin{equation}
U_{\boldsymbol k}(t,t_i)
=
\mathcal{T}
\textrm{exp}
\bigg\{
-\frac{i}{\hbar}
\int_{t_i}^{t} ds H_{\boldsymbol k}(s) 
\bigg\}
,
\label{propagator_analytical}
\end{equation}
where $\mathcal{T}$ indicates time-ordering. Numerically, the propagator $U_{\boldsymbol k}(t,t_i)$ can be evaluated by splitting the time-evolution into $N_T$ sufficiently small time steps of duration $\Delta t$, namely, 
\begin{equation}
U_{\boldsymbol k}(t,t_i)
\approx 
\mathcal{T} 
\prod_{N=1}^{N_T} 
\textrm{exp} 
\bigg\{
-\frac{i}{\hbar}
\Delta t 
H_{\boldsymbol k}( t_N ) 
\bigg\}
\label{propagator_numerical}
\end{equation}
with 
\begin{equation}
t_N = t_i + (N - 1) \Delta t
,
\label{time_grid}
\end{equation}
\begin{equation}
t - t_i = N_T \Delta t
.
\label{total_time}
\end{equation}
The question we want to answer is: How close is the density operator $\rho_{\boldsymbol k}(t_i + \tau)$, resulting from the exact quantum dynamics under the amplitude modulated control, to what one would ideally expect from Floquet theory? In particular, the \textit{targeted} density operator is
\begin{equation}
\rho_{\boldsymbol k}^{(\text{T})}(t_i + \tau)
=
\sum_{n}^{}
f'_{n \boldsymbol k}\,
\Phi_{n \boldsymbol k}(t_i + \tau)
\Phi^{\dagger}_{n \boldsymbol k}(t_i + \tau)
,
\label{targeted_density_matrix}
\end{equation}
indicating that the state we desire to achieve consists of the Floquet modes each having the occupation $f'_{n \boldsymbol k}$. 

The quantum fidelity is a rigorous measure for comparing the overlap between two density operators \cite{nielsen2010quantum},
\begin{equation}
\mathcal{F}_{\boldsymbol k}
\equiv 
\frac{1}{N_p^2}
\bigg[
\text{Tr}
\bigg(
\sqrt{ \rho_{\boldsymbol k}^{(\text{T})} }
\rho_{\boldsymbol k}
\sqrt{ \rho_{\boldsymbol k}^{(\text{T})} }
\bigg)^{\!1/2\,}
\bigg]^2
.
\label{fidelity_formula}
\end{equation}
For this fidelity metric, we have $\mathcal{F}_{\boldsymbol k} = 0$ if there is no overlap, while $\mathcal{F}_{\boldsymbol k} = 1$ signifies that the state has been prepared perfectly. For concision the time variable is suppressed in Eq.~\eqref{fidelity_formula}. In particular, the fidelity will be evaluated at the end of the ramping, i.e., at $t = t_i + \tau$.

The previous section established the energy band structures and eigenstates of the static and the Floquet driven system. Taking that information as a reference point, we may now study the exact quantum dynamics (Eqs.~\eqref{td_ham_general} and~\eqref{ramp_general}) of the more realistic case where the external control is modulated on from an initial amplitude of $V_0 = 0$ to a targeted amplitude of $V_0 = V_T$. While our formalism is general and arbitrary states can be studied, we take the initial and targeted density operators (see Eqs.~\eqref{initial_density_matrix} and~\eqref{targeted_density_matrix}) to be pure states of the form
\begin{equation}
\rho_{\boldsymbol k}(t_i)
=
\varphi_{1 \boldsymbol k}
\varphi^{\dagger}_{1 \boldsymbol k}
,
\label{initial_density_matrix_qw}
\end{equation}
\begin{equation}
\rho_{\boldsymbol k}^{(\text{T})}(t_i + \tau)
=
\Phi_{1 \boldsymbol k}(t_i + \tau)
\Phi^{\dagger}_{1 \boldsymbol k}(t_i + \tau)
,
\label{targeted_density_matrix_qw}
\end{equation}
which both correspond to the lower band being fully occupied for all $\boldsymbol k$.


\subsection{Elementary Monotonic Ramps}
\label{sec:monotonic}

A natural first step is to take the ramp $R(t)$ to be described in terms of elementary monotonic functions such as polynomials, exponentials, logarithms, or sinusoids. Concretely, the following forms are considered: 
\begin{equation}
R(t)
=
\bigg(
\frac{t - t_i}{\tau}
\bigg)^\mathcal{P}
,
\hspace{0.25cm}
\mathcal{P} > 0
,
\label{polynomial_ramp}
\end{equation}
\begin{equation}
R(t)
=
\frac{1}{\mathcal{P} - 1}
\Big[
\mathcal{P}^{(t - t_i)/\tau}
-
1
\Big]
,
\hspace{0.25cm}
\mathcal{P} > 1
,
\label{exponential_ramp}
\end{equation}
\begin{equation}
R(t)
=
\bigg(
\frac{1}{\textrm{log}_\mathcal{P}2}
\bigg)
\textrm{log}_\mathcal{P} 
\bigg[
1 
+ 
\bigg(
\frac{t - t_i}{\tau}
\bigg) 
\bigg]
,
\hspace{0.25cm}
\mathcal{P} > 1
,
\label{logarithmic_ramp}
\end{equation}
\begin{equation}
R(t)
=
\textrm{sin}^{\mathcal{P}}
\bigg[
\frac{\pi}{2}
\bigg(
\frac{t - t_i}{\tau}
\bigg)
\bigg]
,
\hspace{0.25cm}
\mathcal{P} > 0
,
\label{sine_ramp}
\end{equation}
with the adjustable power parameter $\mathcal{P}$ in each case.

\begin{figure}[!htb]
\centering{\includegraphics[scale=0.55]{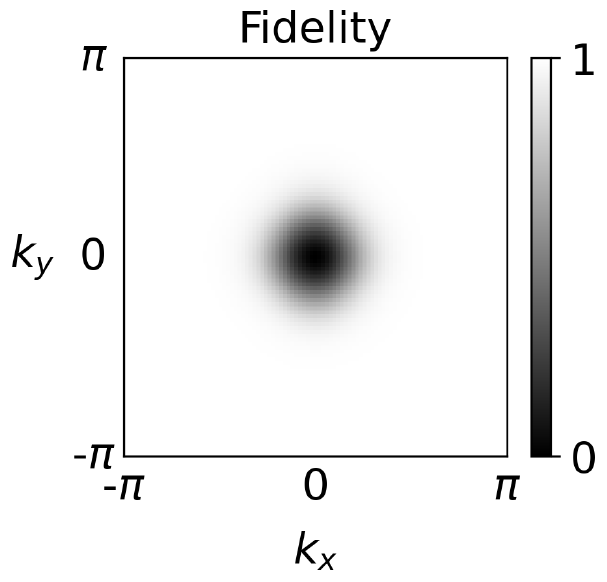}}
\caption{
{\bf Fidelity as a function of crystal momentum.} A linear ramp of 10 Floquet driving cycles in duration is used [Eq.~\eqref{polynomial_ramp}, with $\mathcal{P} = 1$]. As expected, the fidelity is poor at $\boldsymbol k = \Gamma$, where the topological gap closing point occurs in the quasienergy band structures from the previous figure.
}
\label{fig:linfid}
\end{figure}

As an example, in Fig.~\ref{fig:linfid} we plot the fidelity (Eq.~\eqref{fidelity_formula}) as a function of $\boldsymbol k$ in the first Brillouin zone for a linear ramp over ten Floquet cycles with $\mathcal{P} = 1$ and $\tau = 10T$ in Eq.~\eqref{polynomial_ramp}. Away from $\Gamma$, it is trivial to realize high fidelities; however, near the $\Gamma$-point, going through the topological energy gap closing point in real time results in poor fidelities, with potential errors in both the Floquet modes themselves and their occupations, by inspection of the targeted density operator (Eq.~\eqref{targeted_density_matrix}). This is a realization of the quantum Kibble-Zurek mechanism, whereby traversal of a quantum critical point at a finite rate unavoidably leads to the formation of excitations \cite{zurek1985cosmological, damski2005simplest, deng2011}. The dynamics may be described as a Landau-Zener process near (but not exactly at) the critical momentum $\Gamma$ \cite{landau1932theorie, zener1932non, damski2005simplest}, since the quasienergy bands at a given $\boldsymbol k$ form an avoided crossing when the drive amplitude is modulated in the vicinity of the critical value $V_0 = V_c$. Focusing on just $\Gamma$, we also compute the fidelity by varying $\tau$ and $\mathcal{P}$ over a representative range of values for the four varieties of elementary ramp. In the best case scenario, the fidelity at $\Gamma$ can reach up to about 0.05 when the ramp approaches a step function, i.e., an instantaneous quench \cite{wilson2016remnant, schmitt2017universal}. Intuitively, instead of varying the amplitude continuously through the critical value $V_0 = V_c$, one gets the best result by jumping directly from $V_0 = 0$ to $V_0 = V_T$.


\subsection{Simple Oscillating Ramps}
\label{sec:oscillating}

Based on the above results, monotonic ramps are not particularly useful for improving the fidelity at the topological energy gap closing point. One possibility is then to introduce oscillations into the ramp. In perhaps the simplest case, one can make a small modification to Eq.~\eqref{sine_ramp}, yielding
\begin{equation}
R(t)
=
\textrm{sin}^2
\bigg[
\mathcal{C}
\frac{\pi}{2}
\bigg(
\frac{t - t_i}{\tau}
\bigg)
\bigg]
,
\hspace{0.25cm}
\mathcal{C} = 1,3,5,\ldots
,
\label{simple_oscillating_ramp}
\end{equation}
where the integer $\mathcal{C}$ 
counts the number of times the ramping function brings the driving term in the time-dependent Hamiltonian through the critical amplitude value that corresponds to the energy gap closing point.

\begin{figure*}[!htb]
\centering{\includegraphics[scale=0.55]{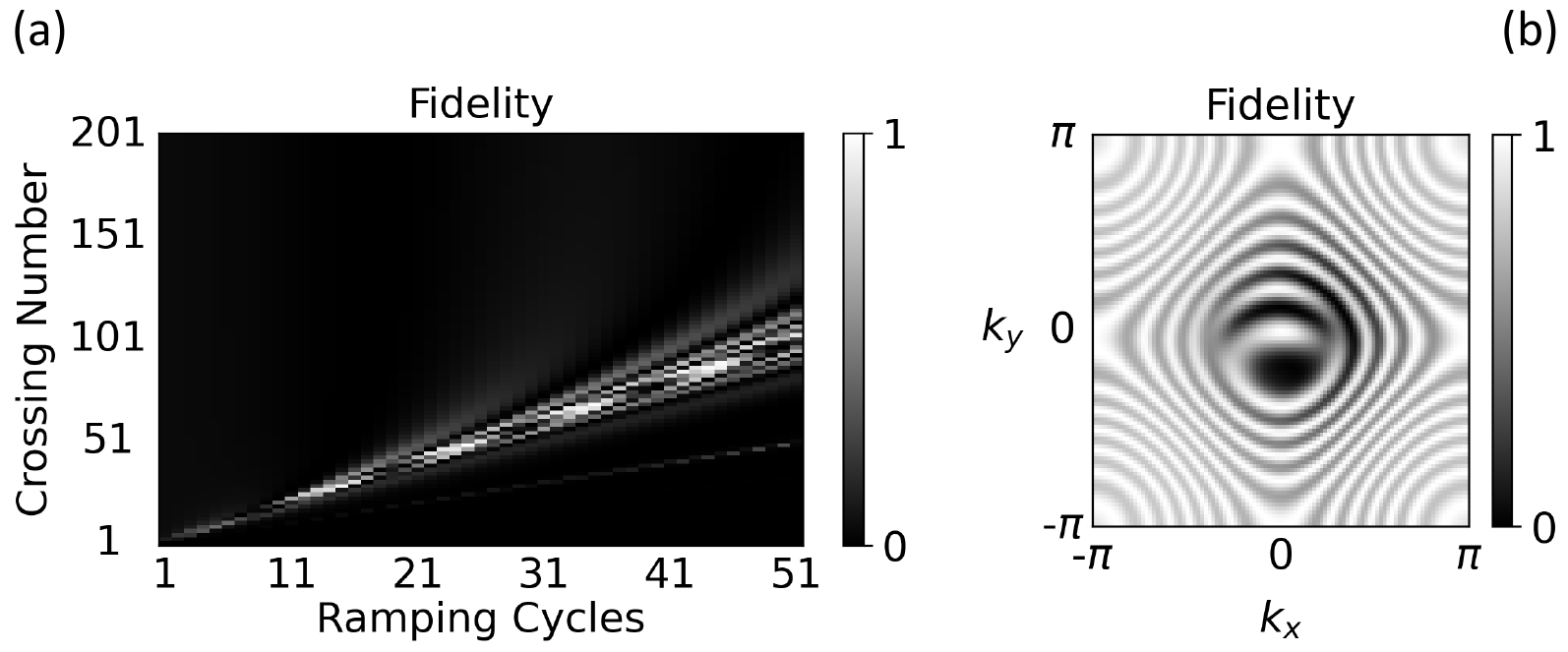}}
\caption{
{\bf Fidelity improvement by parameter optimization of a simple oscillating ramp} 
[Eq.~\eqref{simple_oscillating_ramp}]. 
(a) Fidelity at $\boldsymbol k = \Gamma$ as a function of the number of Floquet ramping cycles $N_R$ and the crossing number $\mathcal{C}$ that counts the number of times the amplitude is brought through the critical value $V_0 = V_c$. 
(b) The absolute maximum fidelity ($> 0.99$) in panel (a) is achieved for $N^*_R = 35$ and $\mathcal{C}^* = 69$. We plot the fidelity as a function of the entire $k$-space for this parameter combination, which now oscillates significantly compared to the previous figure, despite the significant improvement at $\boldsymbol k = \Gamma$.
} 
\label{fig:oscfid}
\end{figure*}

By varying both the number of ramping cycles $N_R$ ($\tau = N_R T$) and the crossing number $\mathcal{C}$, we construct the $\Gamma$ fidelity map shown in Fig.~\ref{fig:oscfid}a. We see that there is now a number of parameter combinations leading to fidelities close to 1, and that they all appear to fall on a line having $\mathcal{C} \approx 2 N_R$. The absolute maximum fidelity ($> 0.99$) in this figure occurs for $N^*_R = 35$ and $\mathcal{C}^* = 69$. The full fidelity map in the $k$-space for these parameters is visualized in Fig.~\ref{fig:oscfid}b. We see that by improving the fidelity at $\Gamma$, it is compromised elsewhere in the $k$-space. There is a fundamental reason for this, which we discuss in the next section. For now, we notice that the fidelity distribution in the $k$-space is not smooth and contains an oscillatory structure. Furthermore, it is instructive to re-write the ramping function Eq.~\eqref{simple_oscillating_ramp} with the optimal parameters as
\begin{equation}
R(t)
=
\frac{1}{2}
\left[
1
-
\textrm{cos} \left( \frac{69}{70}\Omega (t -t_i) \right)
\right]
.
\label{ramp_35_69}
\end{equation}
Multiplying this by $\textrm{cos}(\Omega t)$ or $\textrm{sin}(\Omega t)$ gives the overall time-dependence of $V(t)$, since $V(t) = R(t) D(t)$. Then, beyond the original frequency component at $\Omega$, $V(t)$ will acquire two additional contributions, at the difference and sum, that is, $(1/70)\Omega$ and $(139/70)\Omega$ in this case, respectively. The first is a slow envelope that modulates the fast driving components at the Floquet frequency $\Omega$ and an additional second component at approximately twice this value. Therefore, in practice, the simple oscillating ramp for this combination of parameters is not really just an amplitude modulation of the pure Floquet drive. Rather, we have a {\em two-tone preparation scheme} \cite{minguzzi2022topological}, with a precise phase relationship between the different components.


\section{Floquet State Preparation \\ by Quantum Optimal Control}
\label{sec:qoc}

In the previous sections, we described what may be called the forward problem, that is, one specifies the ramping function, simulates the time-evolution of the density operator, and then computes the fidelity. What we really want is to solve the inverse problem, however: If the fidelity is a functional of the ramping function, what is the ramping function that maximizes the fidelity, subject to the appropriate dynamical and boundary constraints? This may be naturally cast in the language of an optimal control problem, where maximizing the fidelity is the relevant control objective. Formally, we have the following variational problem for a given $\boldsymbol k$:
\begin{equation}
\frac{ \delta \mathcal{F}_{\boldsymbol k}[R(t)] }{ \delta R(t) } = 0
, 
\label{fidelity_variational_problem}
\end{equation}
which can be tackled effectively with QOC methods \cite{boscain2021introduction, ansel2024introduction, duncan2025taming, lambert2026qutip, goerz2019krotov}. The intersection of Floquet physics and QOC theory has been explored in a number of different contexts in the literature \cite{bartels2013smooth, castro2022floquet, castro2023floquet, castro2023optimizing, castro2024qocttools, arrouas2023floquet, castro2025floquet}. For the state preparation problem in particular, we utilize the gradient ascent in function space (GRAFS) method \cite{lucarelli2018quantum}, whereby the ramping function is expanded in a linear combination of basis functions
\begin{equation}
R(t)
=
c_0
+
\sum_{b = 1}^{N_b}
c_b R_b(t)
.
\label{ramp_expansion_basis_functions}
\end{equation}
In this way, the variational problem in Eq.~\eqref{fidelity_variational_problem} is recast as a multivariable optimization problem with respect to the expansion coefficients $c_B$,
\begin{equation}
\frac{ \partial \mathcal{F}_{\boldsymbol k} }{ \partial c_B } = 0, \quad \forall B
.
\label{fidelity_multivariable_optimization_problem}
\end{equation} 
We choose a Fourier basis to easily introduce simple oscillations \cite{castro2022floquet, pagano2024role}, that is, 
\begin{equation}
R(t)
=
c_0
+
\sum_{b=1}^{N_b}
c_b\,
\textrm{cos}[b \tilde{\Omega} (t - t_i)]
.
\label{ramp_fourier_basis}
\end{equation}
Specifically, we use the even Fourier series around the point $t = t_i$, with a period of $\tilde{T} = 2 \tau$ and a corresponding fundamental angular frequency of $\tilde{\Omega} = 2 \pi / \tilde{T}$ so that $R(t_i-\tau) = R(t_i + \tau) = 1$. The benefit of the GRAFS QOC method in this context is that the frequency composition of the ramp is discrete and the maximum frequency component is limited by finding the minimum value of $N_b$ that achieves the desired fidelity. From an experimental perspective, it is easier to generate a pump that combines phase-coherent harmonics of the same fundamental frequency. We note that in this work we generalize previous approaches \cite{lucarelli2018quantum, castro2022floquet} from wave functions to general density operators.

\begin{figure*}[!htb]
\centering{\includegraphics[scale=0.5]{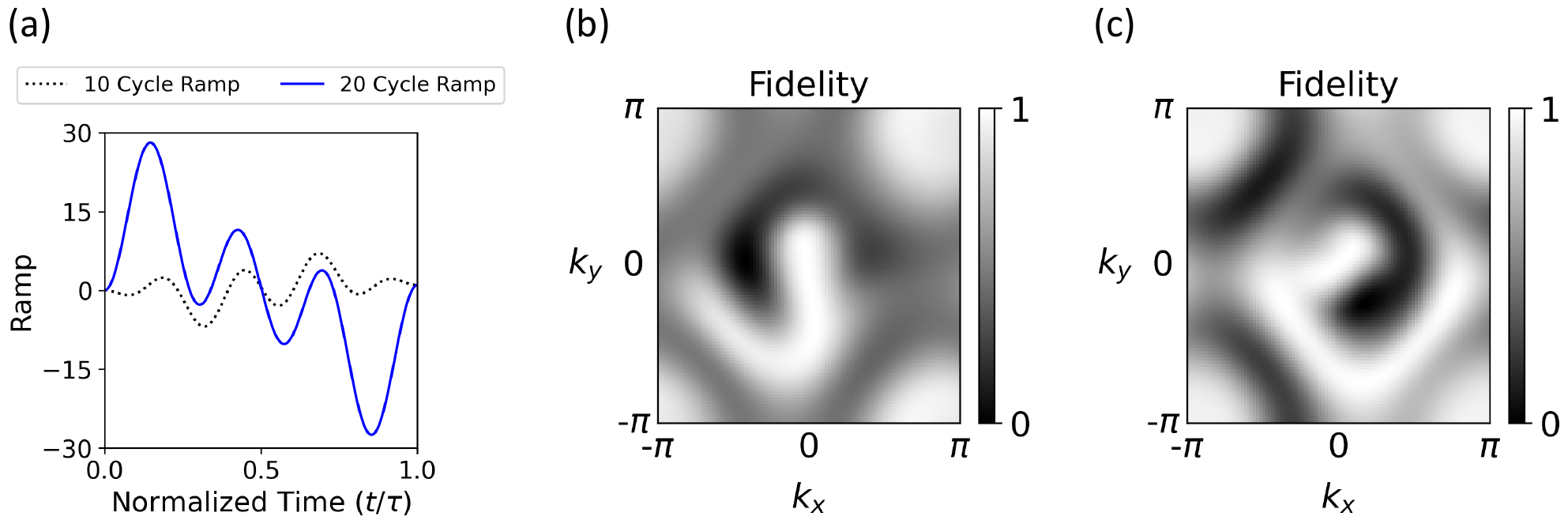}}
\caption{
{\bf Fidelity results for optimally designed ramps.} 
(a) Ramping profiles optimized by QOC with the requirement that the fidelity at $\boldsymbol k = \Gamma$ be improved to better than 0.99. The maximum frequency component in these solutions is considerably lower than that of the optimized simple oscillating ramp in the previous figure. 
(b) Corresponding fidelity map in the $k$-space for the 10 Floquet ramping cycle optimized ramp. 
(c) Same as panel (b), except for 20 Floquet ramping cycles. The fidelity is now a much smoother function of $\boldsymbol k$ overall.
} 
\label{fig:qocrampfid}
\end{figure*}

While the simple oscillating ramp of the previous section can improve the fidelity significantly at $\Gamma$, ideally one would like to find a ramp that is as-simple-as-possible and produces an increased fidelity, but also a smoothed distribution near $\Gamma$ in the $k$-space. This can be addressed by solving the inverse problem defined by Eq.~\eqref{fidelity_variational_problem} at $\boldsymbol k = \Gamma$, using the GRAFS QOC method. We focus on ramp times of $\tau = 10T$ and $\tau = 20T$. In each case, we begin with $N_b = 3$ for the number of basis coefficients in Eq.~\eqref{ramp_fourier_basis}. For each of 50 random initializations of the set of basis coefficients $\{ c_B \}$, we utilize the results of Appendix~\ref{sec:app-ders} to compute the required sets of fidelity derivatives $\{ \partial \mathcal{F}_\Gamma / \partial c_B \}$, in order to perform a gradient ascent optimization of $\mathcal{F}_\Gamma$, as described in Appendix~\ref{sec:app-pga}. Since each optimization finds a local maximum of the multivariable fidelity surface, performing many trials enables an attempt at finding the global maximum. If none of the 50 trials lead to $\mathcal{F}_\Gamma > 0.99$, then we increase $N_b$ by 1 and repeat the procedure. For $\tau = 10T$ and $\tau = 20T$, we require $N_b = 9$ and $N_b = 7$, respectively. In each case, since there are several valid solutions, we choose the simplest one, in a sense we now clarify. Mathematically, we consider this to be the trial with the lowest average frequency defined as
\begin{equation}
\Omega_\textrm{avg}
\equiv
\sum_{b=1}^{N_b} b \tilde{\Omega} |c'_b|
,
\label{avg_frequency}
\end{equation}
where the $c_b$ are all rescaled by the same factor to produce the $c'_b$ so that $\sum_{b=1}^{N_b} |c'_b| = 1$, thereby yielding a normalized discrete probability distribution.

The optimized ramps for $\tau = 10T$ and $\tau = 20T$ are visualized in Fig.~\ref{fig:qocrampfid}a, and the corresponding fidelity maps in the $k$-space are provided in panels (b) and (c) of the same figure, respectively. There are two main benefits to the optimally designed ramps: 

(i) The ratio of the maximum frequency in the ramp based on Eq.~\eqref{ramp_fourier_basis} to the Floquet frequency, $\Omega_\textrm{max}/\Omega = b_\textrm{max} / 2 N_R$, is significantly reduced compared to the optimized simple oscillating ramp of Sec.~\ref{sec:oscillating}. In particular, for $\tau = 10T$ and $\tau = 20T$, we have $\Omega_\textrm{max}/\Omega = 9/20 = 0.45$ and $\Omega_\textrm{max}/\Omega = 7/40 = 0.175$, respectively. This is to be contrasted with $\Omega_\textrm{max}/\Omega = 69/70 \approx 0.986$ for the optimized simple oscillating ramp with $\tau^* = 35T$ and $\mathcal{C}^* = 69$. 

(ii) The optimal ramps create a much larger region of enhanced fidelity near $\Gamma$ in the $k$-space, with distributions being much smoother overall, possibly as a consequence of the previous point. However, a potential drawback stems from the fact that the magnitude of the optimal ramps can extend reasonably beyond 1 now. So long as such short-lived increases in the strength of the time-dependent potential do not damage the material, this is not an issue in practice. This would need to be assessed on a case-by-case basis. 

Physically, the optimal oscillation of the driving amplitude may be thought to lead to a maximal {\em coherent cancellation} of diabatic errors. On the one hand, this has analogies with coherent averaging of interactions by rapid control modulation, as in dynamical decoupling methods \cite{Haeberlen, viola1999dynamical}. On the other hand, this is also in line with the dynamical cancellation of excitations that has been found to arise from interference effects in repeated traversals through a quantum critical point -- in the context of Kibble-Zurek non-equilibrium scaling and Landau-Zener(-St\"uckelberg) physics \cite{kayanuma1993phase, mukherjee2009effects, shevchenko2010landau, berdanier2017floquet, kou2022interferometry, ivakhnenko2023nonadiabatic}. 

Lastly, we see from the fidelity distributions in the $k$-space that when the fidelity is improved at $\Gamma$, it is compromised elsewhere. Fundamentally, this results from the fact that the Chern number is provably conserved in time under arbitrary unitary dynamics \cite{d2015dynamical}. If the fidelity was engineered to 1 at every $\boldsymbol k$, the Chern number would in fact change from 0 to 1, which contradicts the stated theorem. Therefore, we can only engineer the fidelity \emph{distribution} in the $k$-space. Future work may investigate what precise limitation the Chern number conservation mandates on the fidelity {\em averaged} over the first Brillouin zone.


\section{Anomalous Hall Transport}
\label{sec:transport}

For all of the ramps under consideration, ultimately one would like to determine the expectation value of a physical observable that is accessible experimentally with current technology. A prominent such example is the time-dependent anomalous Hall current \cite{mciver2020light}, on which we focus next.

\begin{figure*}[!htb]
\centering{\includegraphics[scale=0.7]{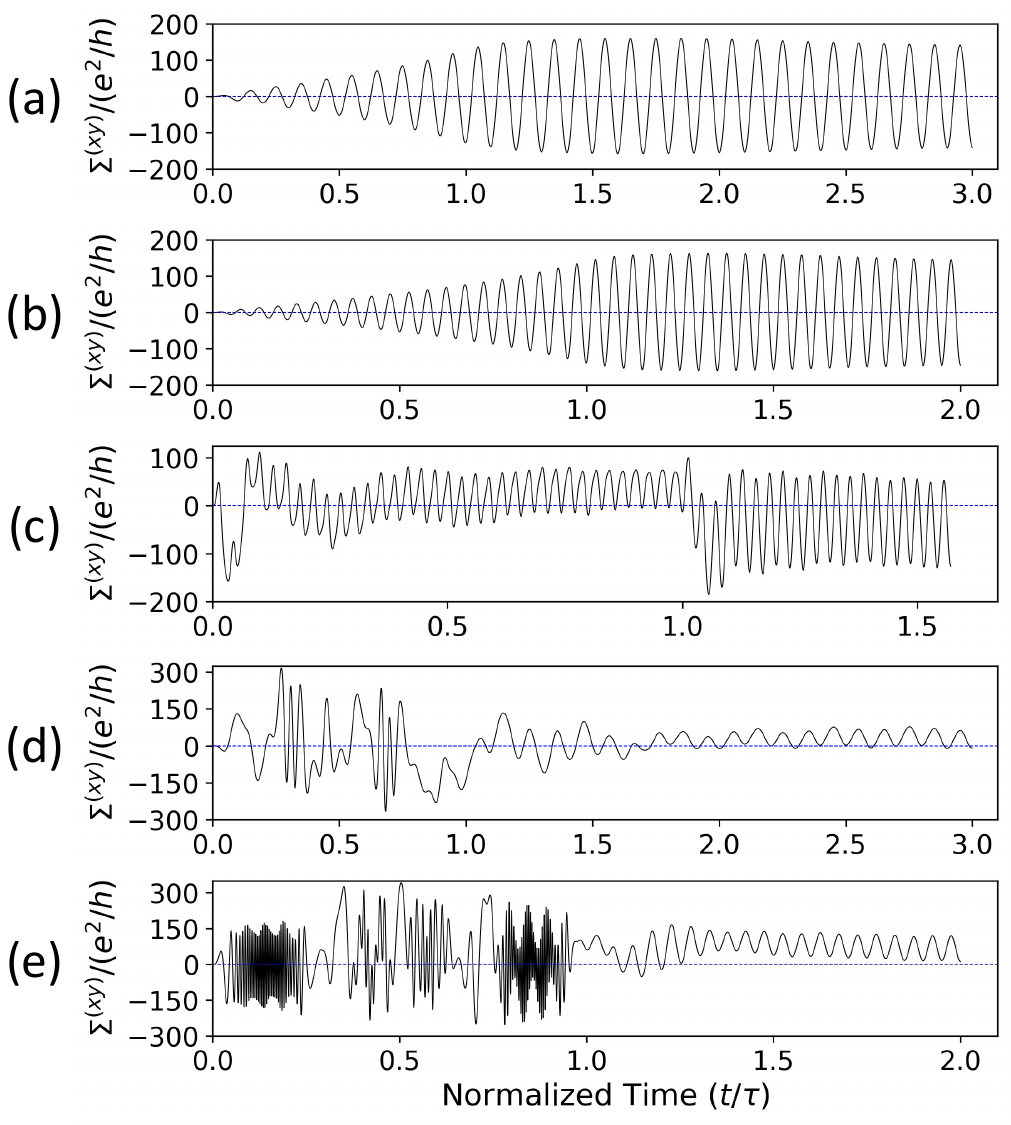}}
\caption{
{\bf Computed time-dependent anomalous Hall conductivities under different ramps.} 
(a) Linear ramp over 10 Floquet driving cycles. 
(b) Linear ramp over 20 cycles. 
(c) Simple oscillating ramp over 35 cycles with a crossing number $\mathcal{C}^* = 69$. 
(d) Optimally designed ramp over 10 cycles. 
(e) Optimally designed ramp over 20 cycles. The oscillating ramps produce conductivity profiles with considerable more complexity.
} 
\label{fig:tdHallcond}
\end{figure*}


\subsection{Hall Transport Theory}
\label{sec:transport_theory}

On top of the amplitude modulated Floquet drive, one now adds a weak probing DC electric field along the $y$ direction and then computes the expectation value of the 2D electronic current density along the $x$ axis \cite{hu2016dynamical, peralta2018time, sato2019microscopic, sato2019light, ge2021universal}. The probing electric field is included in the Hamiltonian as a magnetic vector potential via minimal coupling, that is, 
\begin{equation}
H_{\boldsymbol k}(t)
\rightarrow
H_{\boldsymbol K(t)}(t)
,
\label{ham_transport}
\end{equation}
\begin{equation}
\boldsymbol k
\rightarrow
\boldsymbol K(t) = \boldsymbol k + \frac{|e|}{\hbar} \boldsymbol A(t)
,
\label{minimal_coupling_weak_probe}
\end{equation}
whereby the relevant vector potential and electric field are, respectively, determined as follows: 
\begin{align}
\boldsymbol A(t)
=
-E_0
\{
&(t - t_i + t_p - \tau_p)
\nonumber
\\
&+
\tau_p \textrm{exp}[-(t - t_i + t_p) / \tau_p]
\}
\hat{y}
,
\label{vector_potential_weak_probe}
\end{align}
\begin{equation}
\boldsymbol E(t) = - \partial_t \boldsymbol A(t)
,
\label{electric_field_weak_probe}
\end{equation}
\begin{equation}
\boldsymbol E(t)
=
E_0
\{
1 - \textrm{exp}[-(t - t_i + t_p) / \tau_p]
\}
\hat{y}
.
\label{electric_field_turn_on}
\end{equation}
Notice that the probing field is slowly turned on in the time range $t_i - t_p$ to $t_i$ prior to the amplitude modulation of the external control, in order to prevent unwanted excitations. Concretely, we take $t_p = 10T$ and set $\tau_p = 2T$ to ensure a smooth turn-on. 

\begin{figure*}[!htb]
\centering{\includegraphics[scale=0.6]{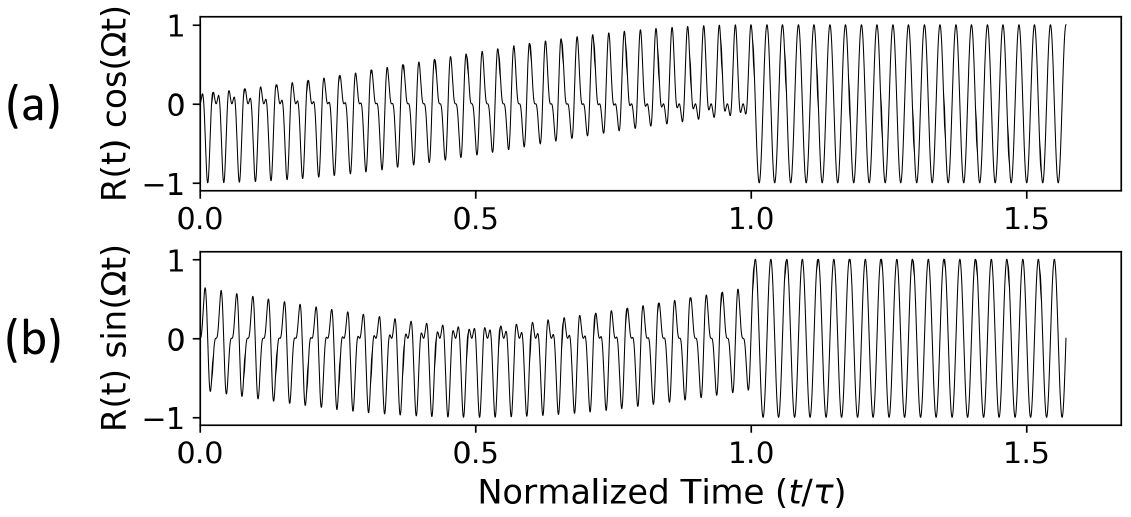}}
\caption{
{\bf Time-dependence of the driving potential resulting from effective two-tone modulation.}
The product of the simple oscillating ramp over $N^*_R = 35$ driving cycles with a crossing number $\mathcal{C}^* = 69$ with the cosine and sine terms from the Floquet drive yields the full time-dependence of $V(t)$. The jumps in amplitude and/or symmetry at $t = 0$ and $t = \tau$ are correlated with the conductivity jumps in panel (c) of the previous figure at the same times.
} 
\label{fig:condjumps}
\end{figure*}

The 2D time-dependent current density operator and its corresponding expectation value are given as
\begin{equation}
J^{(x)}_{\boldsymbol K(t)}(t)
=
-\frac{\partial H_{\boldsymbol K(t)}(t)}{\partial A_x}
,
\label{current_density_operator}
\end{equation}
and, respectively,
\begin{equation}
\langle J^{(x)} \rangle(t)
=
\int_{\textrm{BZ}}^{} \frac{d^2 k}{(2 \pi)^2}
\textrm{Tr}
\left\{
\rho_{\boldsymbol K(t)}(t)
J^{(x)}_{\boldsymbol K(t)}(t)
\right\}
.
\label{current_density_expectation_value}
\end{equation}
For our particular problem, we have
\begin{equation}
J^{(x)}_{\boldsymbol K(t)}(t)
=
-\frac{|e|}{\hbar} l_x
[
A \textrm{cos}(k_x l_x) \sigma_x - 2 B \textrm{sin}(k_x l_x) \sigma_z
]
,
\label{specific_current_density_operator}
\end{equation}
which is time-independent since the probe is along $y$ while the current is along $x$. Similar to Eqs.~\eqref{lvn_equation}-\eqref{propagator_analytical}, the density operator $\rho_{\boldsymbol K(t)}(t)$ satisfies
\begin{equation}
i \hbar \partial_t \rho_{\boldsymbol K(t)}(t) 
= 
[ H_{\boldsymbol K(t)}(t) , \rho_{\boldsymbol K(t)}(t) ]
\label{lvn_equation_transport}
\end{equation}
with a formal solution \footnote{For the transport simulations we take $\rho_{\boldsymbol k}(t_i-t_p)$ to be of the same form as $\rho_{\boldsymbol k}(t_i)$ from Eq.~\eqref{initial_density_matrix}.}
\begin{equation}
\rho_{\boldsymbol K(t)}(t)
=
\tilde{U}_{\boldsymbol K(t)}(t,t_i-t_p)
\rho_{\boldsymbol k}(t_i-t_p)
\tilde{U}^{\dagger}_{\boldsymbol K(t)}(t,t_i-t_p)
,
\label{lvn_solution_transport}
\end{equation}
\begin{equation}
\tilde{U}_{\boldsymbol K(t)}(t,t_i-t_p)
=
\mathcal{T}
\textrm{exp}
\bigg\{
-\frac{i}{\hbar}
\int_{t_i-t_p}^{t} ds H_{\boldsymbol K(s)}(s) 
\bigg\}
.
\label{propagator_transport}
\end{equation}
For the transport simulations, after the ramping is complete at time $t = t_i + \tau$, we continue driving the system periodically for a total time of $20T$. For a sufficiently weak probe amplitude $E_0$, the time-dependent anomalous Hall conductivity is defined by \cite{hu2016dynamical}
\begin{equation}
\Sigma^{(xy)}(t)
\equiv
\frac{\langle J^{(x)} \rangle(t)}
{E_y(t)}
,
\label{td_conductivity}
\end{equation}
where $E_y(t)$ is the $y$ component of the probing electric field in Eq.~\eqref{electric_field_turn_on}.


\subsection{Generation of Ultrahigh \\ Anomalous Hall Conductivities}
\label{sec:transport_results}

Following the above prescription, we compute the time-dependent anomalous Hall conductivity (Eq.~\eqref{td_conductivity}) using a weak probe amplitude of $E_0 = 10^{-3}$ for several ramp profiles. For reference, the first two are linear ramps with $\tau = 10T$ and $\tau = 20T$ (Eq.~\eqref{polynomial_ramp} with $\mathcal{P} = 1$). The third is the optimized simple oscillating ramp with $\tau^* = 35T$ and $\mathcal{C}^* = 69$ (Eq.~\eqref{simple_oscillating_ramp}). The last two are the $\tau = 10T$ and $\tau = 20T$ optimally designed ramps shown in Fig.~\ref{fig:qocrampfid}a. 

The results of the numerical simulations are summarized in Fig.~\ref{fig:tdHallcond}. As expected, in panels (a) and (b) the conductivity for the linear ramps feature simple oscillations with a monotonically increasing amplitude that stabilizes shortly after $t = t_i + \tau$. Although the conductivity oscillations are still simple for the oscillating ramp, we see in panel (c) that there are sudden jumps near $t = t_i$ and $t = t_i + \tau$. This behavior can be qualitatively understood by inspecting the time-dependence of $V(t)$ in the Hamiltonian (Eq.~\eqref{td_ham_general}): In Fig.~\ref{fig:condjumps} we plot the product of the ramp with either $\textrm{cos}(\Omega t)$ or $\textrm{sin}(\Omega t)$. We see that in both cases there are amplitude and/or symmetry changes in the potential's time-dependence precisely at $t = t_i$ and $t = t_i + \tau$, which are correlated with the observed jumps in conductivity. Going back to Fig.~\ref{fig:tdHallcond}, the conductivity for the optimally designed ramps is plotted in panels (d) and (e) for $\tau = 10T$ and $\tau = 20T$, respectively. Although the highest frequency components of these ramps are much lower than that of the simple oscillating ramp, we see that their overall more complex structure and higher amplitudes lead to more complicated dynamics of the charge transport during the turn-on phase. After $t = t_i + \tau$, the conductivity oscillations begin to stabilize around a constant non-zero value. For the optimized parameterization of the simple oscillating ramp in particular, we will see momentarily that this value is approximately half of the crossing number $\mathcal{C}^* = 69$.

\begin{table}[!b]
\centering
\caption{\label{tab:avgHallcond}Time-averaged anomalous Hall conductivities after the ramp is completed.}
\begin{tabular}{cc}
\toprule
Simulation Type & \hspace{3em}$\bar{\Sigma}^{(xy)}/(e^2/h)$ \\
\midrule
(a) Linear Ramp 10 Cycles & \hspace{3em}1.0 \\
(b) Linear Ramp 20 Cycles & \hspace{3em}1.0 \\
(c) Simple Oscillating Ramp 35 Cycles & \hspace{3em}-34.0 \\
(d) Optimal Control Ramp 10 Cycles & \hspace{3em}20.3 \\
(e) Optimal Control Ramp 20 Cycles & \hspace{3em}72.9 \\
\bottomrule
\end{tabular}
\end{table}

\begin{figure*}[!htb]
\centering{\includegraphics[scale=0.6]{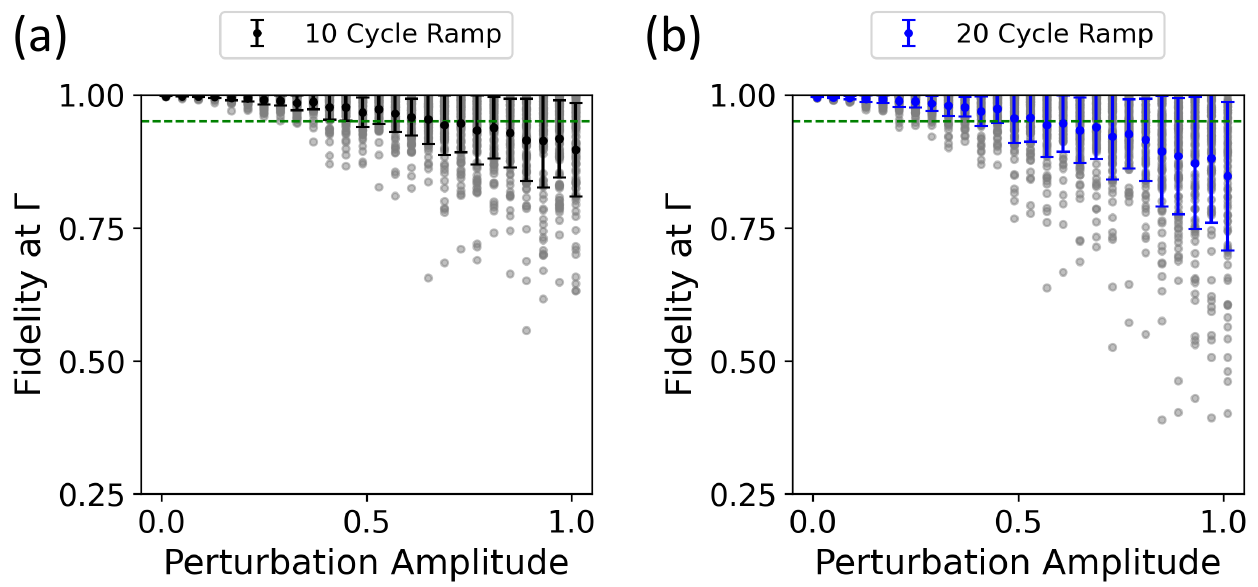}}
\caption{
{\bf Robustness of the fidelity at $\boldsymbol k = \Gamma$ when the optimally designed ramps are perturbed by random noise.} 
At every point on the discrete time grid ($0 < t < \tau$) the ramp is shifted by a number randomly sampled from a Gaussian distribution centered at zero and with standard deviation equal to the perturbation amplitude. To prevent significant outliers, the distribution is truncated at plus or minus three standard deviations. For each perturbation amplitude, 100 independent trials are executed, with the average value indicated by the dark point and the standard deviation of the data set given by the error bar. The horizontal dashed green line is at a value of 0.95 for reference.
} 
\label{fig:robustnessfid}
\end{figure*}

Additionally, it is instructive to time-average the conductivity over an integral number of cycles from $t = t_i + \tau$ to $t = t_i + \tau + 20T$, which has been explored previously both theoretically \cite{sato2019microscopic, sato2019light} and experimentally \cite{mciver2020light}: The resulting values are recorded in Tab.~\ref{tab:avgHallcond}. For the linear ramps, the time-averaged conductivity agrees with what one would expect from the Floquet-generalized TKNN relation, as we have discussed in Sec.~\ref{sec:qw}. This finding is consistent with several previous works that considered time-dependent anomalous Hall transport as a system parameter was modulated monotonically through a topological transition point \cite{hu2016dynamical, sato2019microscopic, sato2019light, ge2021universal}. In stark contrast, the time-averaged anomalous Hall conductivity values for the oscillating ramps can reach \emph{many tens of times what one would naively expect from TKNN}. We believe this is the essential finding of our work. The oscillatory protocol functions as a {\em non-adiabatic topological pump} in 2D \cite{gritsev2012dynamical, titum2016anomalous, privitera2018nonadiabatic, kolodrubetz2018topological, minguzzi2022topological, citro2023thouless}: Repeated traversal through the topological transition leads to the accumulation of time-averaged anomalous Hall current. Physically, we attribute this to a combined effect where, as a result of these repeated forward and backward passages, excitation due to diabatic errors is suppressed, whereas geometric phases from each traversal accumulate constructively.

While a complete theoretical analysis in support of the above interpretation is beyond our current scope, we now elaborate on the implications of this key result. First, a breakdown of the Floquet TKNN relationship (Eq.~\eqref{floquet_tknn}) is evident, indicating that the steady state anomalous Hall conductivity is no longer limited by the Chern number. Practically, the significantly enhanced contrast between the equilibrium and non-equilibrium time-averaged response should enable much easier detection experimentally. On top of this, increasing the conductivity by many tens of times may enable persistence of the phenomena against elevated temperature, scattering in a many-body system, and/or disorder. From these two points follows an improved feasibility of realizing ultrafast topology-based quantum sensing and switching applications within the context of Floquet physics. Moreover, we have shown that the path taken towards the targeted (Floquet) Hamiltonian has a significant impact on physical observables such as anomalous Hall transport, opening many new doors for future fundamental and applied explorations.


\section{Robustness of the Optimally Designed Ramps against Random Noise}
\label{sec:robustness}

\begin{figure*}[!htb]
\centering{\includegraphics[scale=0.6]{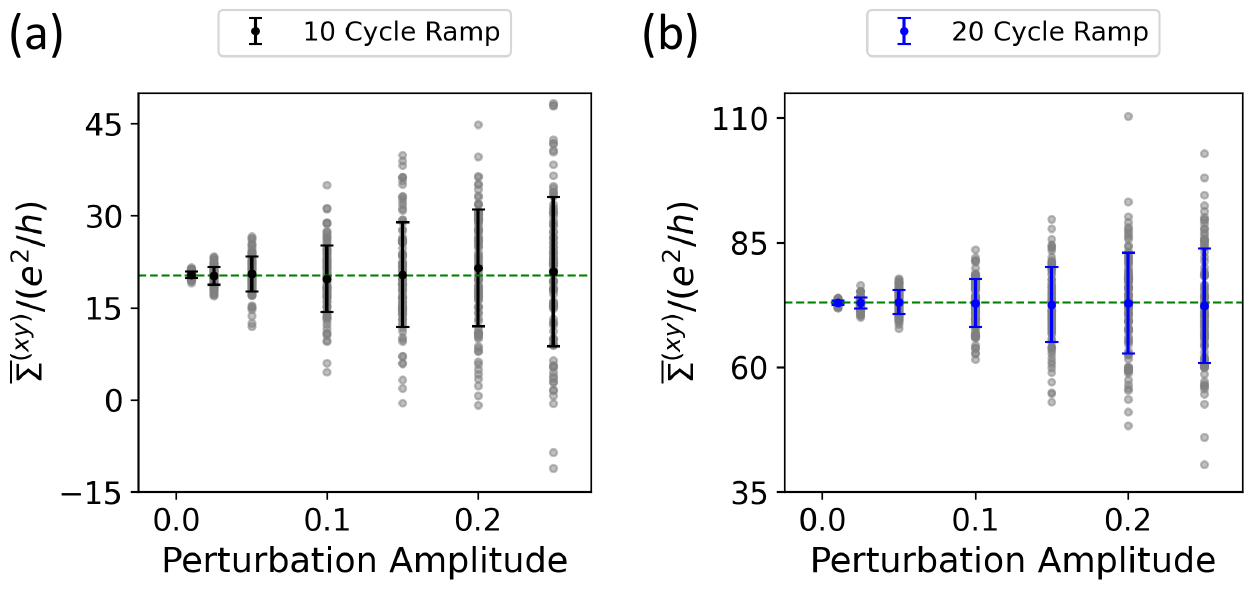}}
\caption{
{\bf Robustness of the post-ramp time-averaged anomalous Hall conductivity against random noise.} 
The horizontal dashed green lines represent the conductivity values from the unperturbed optimally designed ramps that are provided in Tab.~\ref{tab:avgHallcond}. We see that the conductivity values averaged over all of the trials coincide with the ideally calculated value.
} 
\label{fig:robustnesscond}
\end{figure*}

Having practical realizations in mind, an important point to consider is the robustness of the optimally designed ramps against leading imperfections. If these ramps cannot be implemented perfectly in practice, how severe an impact does this have on quantities of interest? In particular, we consider the robustness of the $\Gamma$-point fidelity and the post-ramp time-averaged anomalous Hall conductivity against random noise. To each value of the ramping function on the discrete time grid in the range $t_i < t < t_i + \tau$, we add a random number which is sampled from a Gaussian distribution of mean zero and truncated at plus or minus three standard deviations to prevent non-physical outliers. In what follows, the standard deviation of this Gaussian distribution is what we refer to as the \textit{perturbation amplitude}. For a given perturbation amplitude, we perform 100 trials and compute the quantity of interest for each trial. 

The results for the $\Gamma$-point fidelity are given in Fig.~\ref{fig:robustnessfid}, while the ones for the post-ramp time-averaged anomalous Hall conductivity are shown in Fig.~\ref{fig:robustnesscond}. The gray points are the raw data for each perturbation amplitude and the dark overlay is the average value of the data set with the error bar representing the standard deviation. The system can tolerate a perturbation amplitude of up to around 0.25, while keeping all fidelity points above a threshold of 0.95 (horizontal dashed green line). At the same perturbation amplitude, the conductivity is very sensitive, with raw values spread over a range of about 60 $e^2/h$, including the possibility of sign reversal for the $\tau = 10T$ case. Reducing to a perturbation amplitude of 0.05 brings the range of values down to about 15 $e^2/h$ in both cases. We do notice, however, that the averaged value for each perturbation amplitude coincides with the conductivity from the unperturbed optimally designed ramps (horizontal dashed green lines). These findings will provide experimentalists with an estimate for how much unavoidable random errors in the preparation pulse envelope will affect comparison with theoretical results.


\section{Conclusions and Outlook}
\label{sec:conc}

In summary, we studied the preparation of topological Floquet quantum states from trivial ones, using the 2D quantum well as an illustrative model. Standard monotonic amplitude modulation profiles (ramps) of the Floquet drive result in poor fidelities limited to 0.05 at $\boldsymbol k = \Gamma$, which is where the topological energy gap closing occurs in the Floquet quasienergy band structures. On the other hand, by optimizing over the parameter space of a simple oscillating sine squared ramp, the fidelity can be improved to better than 0.99 at $\Gamma$, while its distribution in the $k$-space becomes very oscillatory. On further analysis, the time-dependent potential turns out to be a slow amplitude modulation of two driving components: one at the Floquet frequency and another at approximately twice this value, with a precise phase relationship between them. On the basis of the gradient ascent in function space (GRAFS) QOC method, we design oscillating ramps for preparation times of 10 and 20 Floquet driving cycles to maximize the fidelity at $\boldsymbol k = \Gamma$ with a minimum cutoff of 0.99. The resulting fidelity distributions in the $k$-space are much smoother. Also, the ratio of the maximum frequency in the ramp to the Floquet frequency is 0.45 and 0.175 for the 10 and 20 cycle cases, respectively, which is significantly reduced compared to the ratio of about 0.986 for the optimized simple oscillating ramp. 

To connect with future experiments, we simulate the time-dependent anomalous Hall transport properties, with a focus on the conductivity values time-averaged an integral number of cycles after the ramping is completed. The linear ramps produce conductivities that correspond to the Chern number as expected from the Floquet generalization of the TKNN relationship. Most importantly, the oscillating ramps all produce what would be considered {\em ultrahigh values} for the 2D quantum well, reaching many tens of times what would be expected if the static system was tuned into a topological regime. We achieve {\em both} a dynamical cancellation of errors and the build-up of time-averaged anomalous Hall current through a dual mechanism: Diabatic errors generated during forward and backward passages interfere destructively to suppress excitation, while geometric phases from each traversal of the topological transition accumulate constructively. Lastly, we study the robustness of the $\Gamma$-point fidelity and the post-ramp time-averaged anomalous Hall conductivities against random noise, in case the optimally designed ramps cannot be implemented perfectly in future experiments. In the process, we have generalized the approaches developed in previous work \cite{lucarelli2018quantum, castro2022floquet} from wave functions to density operators towards broader applicability. 

We conclude with several ideas for future work. Perhaps most importantly, an essential step will be an experimental demonstration of our key finding by extending existing techniques \cite{mciver2020light}. To that end, while we have taken the 2D quantum well model for illustration purposes, our approach is directly applicable to other condensed-matter systems such as graphene \cite{oka2009photovoltaic, sato2019microscopic, mciver2020light}, ultracold atoms in optical lattices \cite{reichl2014floquet, wintersperger2020realization, zhang2023tuning}, and could even be generalized to 3D. Correlating transport measurements \cite{mciver2020light} with sub-cycle resolved photoemission spectra \cite{ito2023build} and topologically protected edge states in the system with open boundary conditions \cite{d2015dynamical} might yield interesting connections. Going beyond the current framework, it is possible that different classes of non-adiabatic oscillatory ramps can be defined that lead to interesting, but fundamentally different modifications to physical observables. In that spirit, instead of focusing on the fidelity itself, one could re-define the QOC problem to maximize the post-ramp time-averaged anomalous Hall conductivity directly \cite{zhang2019floquet}. 

Although we have focused on designing optimal ramps for 10 and 20 Floquet cycle preparation time intervals, perhaps this can be reduced in accordance with known quantum speed limits \cite{caneva2009optimal, deffner2017quantum, bukov2019geometric}. Additionally, the current optimization algorithm puts no restriction on the maximum allowed amplitude in the ramping function, so future implementations could impose constraints on this feature. However, the penalty for reduced amplitudes could very well be increased demands in the frequency domain. The frequency and amplitude characteristics of the external control may also require going beyond a two-band model in general, but this will be specific to the system of interest. While we focus on the $\Gamma$-point in our optimal control calculations, one could instead enhance a composite fidelity containing weighted contributions from multiple crystal momentum vectors in the first Brillouin zone \cite{castro2022floquet}. Lastly, we acknowledged that, due to the Chern number conservation theorem for unitary 2D spatially periodic quantum systems \cite{d2015dynamical}, we can only engineer the fidelity distribution in the $k$-space and not improve it everywhere. An intriguing possibility would be to determine whether the use of non-unitary dynamics, via dissipation engineering \cite{seetharam2015controlled, bandyopadhyay2020dissipative, schnell2024dissipative, ritter2025autonomous}, may enable in principle the fidelity to be improved to 1 everywhere in the $k$-space and, if so, how to achieve this under realistic constraints.


\section*{Acknowledgments}

It is a pleasure to thank Joshuah Heath, Dennis Lucarelli, Shuanglong Liu, Xiao Chen (Northeastern University), Alberto de la Torre, Gregory Fiete, Nuh Gedik, and Anatoli Polkovnikov for stimulating discussions. AC is particularly grateful to Joshuah Heath for many in-depth discussions about the preparation of Floquet states during the early stages of this work. 
Work at Dartmouth College was supported by the NSF under grant No.\,OIA-1921199 and the ARO through US MURI Grant No.\,W911NF1810218. The computations in this work were performed on the Discovery cluster and HPC environments supported by the Research Computing group, IT\&C at Dartmouth College. 
Work at Northeastern University was supported as part of the Center for Molecular Magnetic Quantum Materials, an Energy Frontier Research Center funded by the U.S. Department of Energy, Office of Science, Basic Energy Sciences under Award No.\,\mbox{DE-SC0019330}. The authors acknowledge UFIT Research Computing for providing computational resources (HiPerGator) and support that have contributed to the research results reported in this publication. This research used resources of the National Energy Research Scientific Computing Center (NERSC), a Department of Energy Office of Science User Facility using NERSC award \mbox{BES-ERCAP0022828}.


\appendix


\onecolumngrid

\section{High-Frequency Effective Hamiltonian for the Floquet Driven 2D Quantum Well}
\label{sec:app-hf}

On the basis of Van Vleck degenerate perturbation theory, the effective Hamiltonian in the high-frequency limit is given as 
$H_{\boldsymbol k}^{(\textrm{eff})}
=
H_{\boldsymbol k, 0}^{(\textrm{eff})}
+
H_{\boldsymbol k, 1}^{(\textrm{eff})}
+
\cdots
$,
where \cite{mikami2016brillouin}
\begin{equation}
H_{\boldsymbol k, 0}^{(\textrm{eff})}
=
H_{\boldsymbol k}^{(0)}
,
\qquad 
H_{\boldsymbol k, 1}^{(\textrm{eff})}
=
\sum_{m \neq 0}^{}
\frac{[H_{\boldsymbol k}^{(-m)},H_{\boldsymbol k}^{(m)}]}{2 m \hbar \Omega}
,
\label{eff_ham_01}
\end{equation}
in terms of the Fourier components of the time-periodic Hamiltonian in Sec.~\ref{sec:qw} (Eqs.~\eqref{td_ham_floquet} and~\eqref{time_periodic}). In particular, we have the following sign convention for the Fourier series:
\begin{equation}
H_{\boldsymbol k}(t) 
=
\sum_{m}^{}
H_{\boldsymbol k}^{(m)}
e^{-i m \Omega t}
.
\label{fourier_series_ham}
\end{equation}
The result for our system (Eqs.~\eqref{ham_static},~\eqref{d_vector}, and~\eqref{floquet_drive}) is
\begin{equation}
H_{\boldsymbol k}^{(\textrm{eff})}
=
H_{0, \boldsymbol k}
+
\delta M \sigma_z
,
\label{eff_ham_qw}
\end{equation}
\begin{equation}
\delta M
=
\frac{-4 V_0^2}{\hbar \Omega}
,
\label{eff_M_shift}
\end{equation}
which indicates an effective renormalization of the $M$ parameter in the static quantum well that is proportional to $V_0^2 / \Omega$. On the other hand, if one had attempted a time-periodic drive of the form $D(t) = V_0 \textrm{cos}(\Omega t) \sigma_z$ instead \cite{kumar2020linear}, we see immediately from Eq.~\eqref{eff_ham_01} that the Floquet drive would have no effect on the static Hamiltonian, at least to first order in the series expansion. We confirmed this via exact numerical simulations as well.


\section{Analytical Expression for Fidelity Derivatives}
\label{sec:app-ders}

The essential and computationally expensive part of the GRAFS method \cite{lucarelli2018quantum} is the computation of the derivatives of the fidelity with respect to the ramp basis coefficients, see Eq.~\eqref{fidelity_multivariable_optimization_problem}. Here we derive an analytical expression for these derivatives that is then evaluated numerically, which is more accurate than using finite differences. In the following, unless written explicitly otherwise, $\rho_{\boldsymbol k}$ and $\rho_{\boldsymbol k}^{(\text{T})}$ are evaluated at time $t = t_i + \tau$. Also, the time-evolution operators are written as $U_{\boldsymbol k} = U_{\boldsymbol k}(t_i + \tau,t_i)$ with the times suppressed. We begin with a simpler formula for the quantum fidelity,
\begin{equation}
\mathcal{F}_{\boldsymbol k}
=
\frac{1}{N_p^2}
\bigg(
\text{Tr}
\bigg\{
\sqrt{ \rho_{\boldsymbol k}^{(\text{T})} \rho_{\boldsymbol k} }
\bigg\}
\bigg)^2
,
\label{fidelity_starting_point}
\end{equation}
which has been shown to be equivalent to Eq.~\eqref{fidelity_formula} \cite{muller2023simplified}. For convenience we define
\begin{equation}
F_{\boldsymbol k}
\equiv
\text{Tr}
\bigg\{
\sqrt{ \rho_{\boldsymbol k}^{(\text{T})} \rho_{\boldsymbol k} }
\bigg\}
\label{F_definition}
\end{equation}
so that
\begin{equation}
\mathcal{F}_{\boldsymbol k}
=
\frac{1}{N_p^2}
F_{\boldsymbol k}^2
.
\label{fidelity_re_im_F}
\end{equation}
The necessary fidelity derivatives are derived by repeated application of the chain rule. To start,
\begin{equation}
\frac{\partial \mathcal{F}_{\boldsymbol k}}{\partial c_B}
=
\frac{2}{N_p^2}
F_{\boldsymbol k}
\frac{\partial F_{\boldsymbol k}}{\partial c_B}
,
\label{fid_der_re_im_F}
\end{equation}
\begin{equation}
\frac{\partial F_{\boldsymbol k}}{\partial c_B}
=
\frac{1}{2}
\text{Tr}
\bigg\{
\sqrt{ \left( \rho_{\boldsymbol k}^{(\text{T})} \rho_{\boldsymbol k} \right)^{(-1)} }
\rho_{\boldsymbol k}^{(\text{T})}
\frac{\partial \rho_{\boldsymbol k}}{\partial c_B}
\bigg\}
.
\label{fid_der_dm}
\end{equation}
Recalling Eq.~\eqref{lvn_solution},
$\rho_{\boldsymbol k}
=
U_{\boldsymbol k}
\rho_{\boldsymbol k}(t_i)
U^{\dagger}_{\boldsymbol k}
$,
we have
\begin{equation}
\frac{\partial \rho_{\boldsymbol k}}{\partial c_B}
=
\frac{\partial U_{\boldsymbol k}}{\partial c_B}
\rho_{\boldsymbol k}(t_i)
U^{\dagger}_{\boldsymbol k}
+
U_{\boldsymbol k}
\rho_{\boldsymbol k}(t_i)
\left( \frac{\partial U_{\boldsymbol k}}{\partial c_B} \right)^\dagger
.
\label{dm_der_exact_U}
\end{equation}
Now we re-write Eq.~\eqref{propagator_numerical} as 
\begin{equation}
U_{\boldsymbol k}
= 
\mathcal{T} 
\prod_{N=1}^{N_T} 
U_{\boldsymbol k}^{(N)}
\label{U_product_breakdown}
\end{equation}
in terms of
\begin{equation}
U_{\boldsymbol k}^{(N)}
\equiv 
\textrm{exp} 
\bigg\{
-\frac{i}{\hbar}
\Delta t 
H_{\boldsymbol k}( t_N ) 
\bigg\}
.
\label{each_U}
\end{equation}
By virtue of the product rule we have
\begin{equation}
\frac{ \partial U_{\boldsymbol k} }{ \partial c_B }
=
\sum_{N=1}^{N_T}
U_{\boldsymbol k}^{(N_T)}
U_{\boldsymbol k}^{(N_T-1)}
\cdots
\frac{ \partial U_{\boldsymbol k}^{(N)} }{ \partial c_B }
\cdots
U_{\boldsymbol k}^{(2)}
U_{\boldsymbol k}^{(1)}
.
\label{U_product_breakdown_der}
\end{equation}
For the next step, we re-write the time-dependent Hamiltonian (Eqs.~\eqref{td_ham_general} and~\eqref{ramp_expansion_basis_functions}) in a more useful form ($t_i \leq t \leq t_i + \tau$)
\begin{equation}
H_{\boldsymbol k}(t) 
=
H_{0, \boldsymbol k}
+
R(t) D(t)
=
H_{0, \boldsymbol k}
+
\left( \sum_{b = 1}^{N_b} c_b R_b(t) \right)
D(t)
,
\label{td_ham_general_ramp_expansion}
\end{equation}
\begin{equation}
H_{\boldsymbol k}(t) 
=
H_{0, \boldsymbol k}
+
\left( \sum_{b \neq B}^{N_b} c_b R_b(t) \right)
D(t)
+
c_B ( R_B(t) D(t) )
=
(
H_{0, \boldsymbol k}
+
\tilde{h}_{B}(t)
)
+
c_B h_B(t)
,
\label{td_ham_general_regroup}
\end{equation}
and define the eigenbasis of $H_{\boldsymbol k}(t_N)$
\begin{equation}
H_{\boldsymbol k}(t_N)
| \lambda_{\boldsymbol k i}^{(N)} \rangle
=
\lambda_{\boldsymbol k i}^{(N)}
| \lambda_{\boldsymbol k i}^{(N)} \rangle
.
\label{td_ham_eig}
\end{equation}
On the basis of Refs. \cite{machnes2011comparing, lucarelli2018quantum}, one can write down the matrix elements of $\partial U_{\boldsymbol{k}}^{(N)} / \partial c_B$ as
\begin{equation}
\left\langle \lambda_{\boldsymbol k i}^{(N)} \left| \frac{ \partial U_{\boldsymbol{k}}^{(N)} }{ \partial c_B } \right| \lambda_{\boldsymbol k j}^{(N)} \right\rangle
=
\begin{cases}
-\frac{i}{\hbar} \Delta t \left\langle \lambda_{\boldsymbol k i}^{(N)} \left| h_B(t_N) \right| \lambda_{\boldsymbol k j}^{(N)} \right\rangle\ \textrm{exp}(-i \Delta t \lambda_{\boldsymbol k i}^{(N)} / \hbar ) & \lambda_{\boldsymbol k i}^{(N)} = \lambda_{\boldsymbol k j}^{(N)} \\
\left\langle \lambda_{\boldsymbol k i}^{(N)} \left| h_B(t_N) \right| \lambda_{\boldsymbol k j}^{(N)} \right\rangle\ \frac{ \textrm{exp}(-i \Delta t \lambda_{\boldsymbol k i}^{(N)} / \hbar ) - \textrm{exp}(-i \Delta t \lambda_{\boldsymbol k j}^{(N)} / \hbar ) }{ \lambda_{\boldsymbol k i}^{(N)} - \lambda_{\boldsymbol k j}^{(N)} } & \lambda_{\boldsymbol k i}^{(N)} \neq \lambda_{\boldsymbol k j}^{(N)}
\end{cases}
.
\label{U_der_mat_el}
\end{equation}
This constitutes all of the required formulas to compute the derivatives $\partial \mathcal{F}_{\boldsymbol k} / \partial c_B$ numerically.


\section{Numerical Optimization: Projected Gradient Ascent}
\label{sec:app-pga}

In this appendix we describe the optimization procedure to find the local maximum of the high-dimensional fidelity surface with respect to a given random initialization of the ramp Fourier basis coefficients. First, a given set of randomly initialized basis coefficients will not automatically yield a ramp that satisfies the required boundary conditions implied by Eq.~\eqref{ramp_general}, namely,
\begin{equation}
R(t_i) = 0
,
\qquad
R(t_i + \tau) = 1
.
\label{ramp_bc}
\end{equation}
With the Fourier basis defined in Eq.~\eqref{ramp_fourier_basis}, we see that
\begin{equation}
c_0
=
\frac{1}{2 \tau} \int_{t_i - \tau}^{t_i + \tau} dt R(t)
\label{c0_formula}
\end{equation}
and then impose
\begin{equation}
c_0
=
\frac{R(t_i) + R(t_i + \tau)}{2}
=
\frac{1}{2}
,
\label{c0_impose_avg_bc}
\end{equation}
which is the average of the ramp endpoints. From here, two sum rules for the basis coefficients can be derived
\begin{equation}
\sum_{b=1,3,5,\ldots}^{N_b} c_b = -\frac{1}{2}
,
\qquad 
\sum_{b=2,4,6,\ldots}^{N_b} c_b = 0
.
\label{sum_rule}
\end{equation}
In practice, to make an invalid set of basis coefficients satisfy the required boundary conditions, these sum rules imply the following transformations
\begin{equation}
c_b
\rightarrow
-\frac{1}{2 S_\textrm{odd}} c_b
,
\hspace{3mm}
c_b = 1,3,5,\ldots 
, 
\qquad
c_b
\rightarrow
c_b - \frac{S_\textrm{even}}{N_\textrm{even}}
,
\hspace{3mm}
c_b = 2,4,6,\ldots
,
\label{coeff_rescale}
\end{equation}
where $S_\textrm{odd}$ and $S_\textrm{even}$ are the sums of the odd and even indexed basis coefficients prior to the transformations, and $N_\textrm{even}$ is the total number of $c_b$ with $b$ even. With the corrected ramp basis coefficients, the Fourier series in Eq.~\eqref{ramp_fourier_basis} is summed to yield the ramp $R(t)$ on the discrete time grid. Via Sec.~\ref{sec:qoc} and Appendix~\ref{sec:app-ders}, the $\Gamma$-point fidelity and set of derivatives with respect to each of the basis coefficients can be computed. The basis coefficients are then updated by gradient ascent
\begin{equation}
c_B \rightarrow c_B + \gamma \frac{\partial \mathcal{F}_{\boldsymbol k}}{\partial c_B}
,
\label{coeff_grad_update}
\end{equation}
where the scale factor $\gamma$ is adjusted at each iteration of the optimization algorithm. One first attempts $\gamma = 1$, recomputes the ramping function, and then calculates the maximum of the absolute value of the difference between the old and the new ramp. If this value is less than a cutoff $\Delta_\textrm{cut}$, the update is complete. Otherwise, $\gamma$ is continually reduced by a factor of 2 until the criterion is satisfied. At that point, \textit{for each iteration}, the basis coefficients are projected back into the valid space of possibilities by means of the transformation in Eq.~\eqref{coeff_rescale} in case the gradient update has modified the required boundary conditions encompassed by Eq.~\eqref{ramp_bc}. Such a \textit{projected gradient ascent} algorithm has been described previously in the context of quantum state tomography \cite{bolduc2017projected}. The cutoff value itself depends on the iteration step $N$ as
\begin{equation}
\Delta_\textrm{cut}(N)
=
\frac{\Delta_\textrm{max} - \Delta_\textrm{min}}{1 + e^{\beta (N - N_\textrm{iter}/2)}}
+
\Delta_\textrm{min}
,
\label{cutoff_iter}
\end{equation}
where the maximum and minimum cutoffs are $\Delta_\textrm{max} = 0.1$ and $\Delta_\textrm{min} = 0.001$, the smearing parameter is $\beta = 0.05$, and the total number of iterations is $N_\textrm{iter} = 250$. In this way, larger overall changes to the ramping function are permitted earlier in each optimization simulation. We implement a custom version of this optimization technique from scratch using Python.


\section{Computational Details}
\label{sec:app-comp}

\begin{itemize}
\item All of the codes used to generate the results for this work were written from scratch in Python, utilizing only the basic built-in modules and packages.
\item We sample the square first Brillouin zone with a discrete grid of 101 $\times$ 101 $k$-points.
\item Unless specified otherwise, each Floquet driving cycle is sampled using 101 time points.
\item Chern numbers are computed numerically by utilizing the FHS algorithm \cite{fukui2005chern} in the context of Eq.~\eqref{chern_number}. 
\item In order to ensure accuracy of the exact quantum dynamics, the time grid sampling should be scaled up for the simple oscillating ramps. The highest frequency in the time-dependent potential is $\Omega_\textrm{highest} = \Omega + (\pi \mathcal{C} / \tau)$. The time grid sampling should then be scaled up by the factor $\Omega_\textrm{highest} / \Omega = 1 + (\pi \mathcal{C} / \Omega \tau )$, rounded up to the nearest integer for each calculation. We acknowledge that parameter combinations such as $N_R = 1$ and $\mathcal{C} = 201$ are not physically realistic, but we can still carry out the numerical simulations.
\item On the basis of the random module in Python, cryptographically secure random numbers for the basis coefficient initializations are generated using system sources via the command random.SystemRandom().uniform(-1,1).
\item The matrix inverse in Eq.~\eqref{fid_der_dm} is evaluated in Python using the Moore-Penrose pseudo-inverse to ensure numerical stability.
\item In order to confirm the reliability of our numerical simulations for the two ramps designed by optimal control, we recompute the fidelity maps in the $k$-space and the time-dependent anomalous Hall conductivities (also the time-averaged values) with approximately five times the time grid sampling.
\end{itemize}

\twocolumngrid


\bibliography{references}


\end{document}